\documentstyle[preprint,aps,here,psfig]{revtex}
\draft
\begin{document}
\tightenlines
%%%%%%%%%%%%%%%%%%%%%%%%%%%%%%%%%%%%%%%%%%%%%%%%%%%%%%%%%%%%%%%%%%%%%%
\title{
Cooperative damping mechanism of the resonance
\\
in the nuclear photoabsorption
}
\author{
M. Hirata, 
K. Ochi
\\
{\it Department of physics, Hiroshima University, Higashi-Hiroshima
739, Japan}\\
and \\
T. Takaki
\\
{\it Onomichi Junior College, Onomichi 722, Japan}}
\maketitle
\begin{abstract}
 We propose a resonance damping mechanism to explain
the disappearance of the peaks around the position of the resonances
higher than the $\Delta$ resonance
in the nuclear photoabsorption.
This phenomenon is understood by taking into account
the cooperative effect of the collision broadening of $\Delta$ and
$N^{*}$,
the pion distortion and the interference in the two-pion photoproduction
processes in the nuclear medium.
\end{abstract}
%                    SECTION                                         %
The total nuclear photoabsorption cross section has been measured
on $^{238}$U\cite{exp1}, Be, C\cite{exp2},
$^{7}$Li, C, Al, Cu, Sn and Pb\cite{exp3}.
The data show that the excitation peaks around the position of
the $D_{13}$(1520) and $F_{15}$(1680) resonances disappear
and above 600 MeV there is a reduction of the absolute value of
the cross section per nucleon compared with the data
for hydrogen\cite{exp4,exp5} and deuteron\cite{exp6}.
These experimental findings have been confirmed by the contemporary data
on the photofission cross section of $^{238}$U\cite{exp7} and
$^{235}$U\cite{exp8} obtained at Mainz up to 800 MeV.

 Some theoretical calculations based on simplified 
models have been attempted to explain the strong damping of the peaks
around
the location of the $D_{13}$ and $F_{15}$ resonances\cite{phe1,phe2}.
In both of attempts, the collision broadening of the resonances
has been assumed to be dominant damping mechanism
and very large resonance widths compared with the free ones
were found to be needed to reproduce the experimental data.
However, the collision broadening has been estimated using the
interaction
of the resonance with the nucleon in Ref. \cite{phe3} where
it has been claimed that such significantly increasing
resonance widths were hardly justified.
These theoretical analyses must miss some important effects other
than the collision broadening.

The purpose of this letter is to discuss a mechanism
for the damping phenomena from a qualitative point of view.
Here we focus only on the $D_{13}$ ($N^{*}$) resonance region.
In order to get an idea what is going on in this energy region,
it is necessary to know the details of the elementary process
in the nuclear photoabsorption.
The dominant reactions on a nucleon over the photon energy range
from 500 to 850 MeV are the one-pion and two-pion 
photoproduction\cite{double-exp1}.
The peak of the $D_{13}$ resonance region shows up clearly by the
combined
effect of these reactions as can be seen in Fig. 1
where 
the cross section of the one-pion photoproduction has a small peak at
around 720 MeV and
the cross section of the two-pion photoproduction starts to grow from
around 400 MeV photon energy and increases up to around 800 MeV.
The $D_{13}$ resonance plays an important role in the total cross
sections of these reactions.
In the one-pion photoproduction, the effect of the $P_{33}(1232)$
resonance
diminishes with increasing energy in this energy region and instead the
$D_{13}$ resonance contributes dominantly to produce the small peak
of the cross section at around 720 MeV.
The latter is due to that the $D_{13}$ resonance has a large
photocoupling,
the $\pi N$ branching fraction of 50-60 \% and a width comparable
to
the $P_{33}(1232)$ resonance.
Its photocoupling is at least twice larger than those of the other
relevant resonances such as $P_{11}(1440)$, $S_{11}(1535)$ and
$P_{33}(1600)
$\cite{PDG}.
In our present approach which will be described later, therefore,
the partial wave amplitudes other than $P_{33}$ and $D_{13}$ are
regarded
as background contributions.
In the two-pion photoproduction, the production processes through the
excitation of a resonance by $\gamma N$ coupling [e.g., see Figs.
2(a) and 2(b)] have themselves small contributions and
the dominant contributions are due to the production processes through
the $\Delta$ Kroll-Ruderman, the $\Delta$ pion-pole and the $\rho$
Kroll-Ruderman terms
[Figs. 2(c), 2(d) and 2(e)].
The interference between the $D_{13}$ resonance term and the Kroll-Ruderman
terms, however, is essential to produce the bump of the cross section
\cite{oset,laget,ochi}.
Thus the contribution of the $D_{13}$ resonance term is important compared
with the other resonances.
We take into account only the $D_{13}$ resonance in the resonances excited
by $\gamma N$ coupling.

For the one-pion photoproduction, the analyses due to the partial wave
expansion had been carried out by several authors \cite{arndt}.
The cross sections for both proton and neutron can be obtained
by using partial wave amplitudes given by them. In our calculation
we use the SM95 amplitudes of Arndt {\it et al.}\cite{arndt}.
In the case of proton only a small peak appears near the resonance
energy of $D_{13}$,
but in the case of neutron the energy dependence of cross section
near the same energy shows a shoulder rather than a peak.
This behavior of the $D_{13}$ resonance is quite different from
that of the $\Delta$ resonance.
Therefore, we can easily make the $D_{13}$ resonance peak
for the one-pion photoproduction vanish by introducing much smaller
width due to the collision broadening than those given in
Refs.\cite{phe1,phe2}.

The two-pion photoproduction has been studied by Tejedor and
Oset\cite{oset},
and Murphy and Laget\cite{laget}.
The models have been successful to predict the
$\gamma p \rightarrow \pi^{+} \pi^{-}p$ cross section.
However, predicted cross sections of the
$\gamma p \rightarrow \pi^{+} \pi^{0}n$ and
$\gamma n \rightarrow \pi^{-} \pi^{0}p$ reactions
are too small compared with the experimental
data\cite{double-exp1,double-exp2}.
The total photoabsorption cross section on a nucleon cannot be
reproduced
by using their models.
For the present purpose we need the model which gives a good description
of the two-pion photoproduction.
Recently
we have proposed a simple model to reproduce the total cross section of
two-pion photoproduction\cite{ochi}.
In this model, $N^{*}$, $\Delta$ and $\rho$ meson are treated
in a separable potential model
which is able to describe the resonance over a wide range of
energy\cite{BL}.
The total photoabsorption cross
sections on both the proton and the neutron target can be reproduced
over the
photon energy range between 400 MeV and 850 MeV by using this
model\cite{ochi}, e.g., see Fig. \ref{fig:GP}.
Hereafter we discuss the two-pion photoproduction based on our model.

The two-pion photoproduction takes place through the $\pi \Delta$
or the $\rho N$ channel.
The $N^{*}$ resonance decays into a $\pi \pi N$ state through these
channels
which diagrams are shown in Figs. \ref{fig:fig1}(a) and
\ref{fig:fig1}(b).
The corresponding $T$ matrices are denoted by
$T_{N^{*}(\pi \Delta)}^{s(d)}$ and $T_{N^{*}\rho N}$ respectively,
where $s(d)$ means a $s$-wave or $d$-wave $\pi \Delta$ state.
In the $\gamma p \rightarrow \pi^{+} \pi^{-} p$ reaction,
the dominant processes are the $\Delta$ Kroll-Ruderman term
($T_{\Delta {\rm KR}}$) in Fig. \ref{fig:fig1}(c) and
$\Delta$ pion-pole term ($T_{\Delta {\rm PP}}$) in Fig.
\ref{fig:fig1}(d).
The $N^{*}$ contribution alone is small but the interference between
$T_{N^{*}(\pi \Delta)}$ and $T_{\Delta {\rm KR}}$ is important.
Because of this, the $N^{*}$ excitation is regarded as an important
ingredient
in the two-pion photoproduction.
In the $\gamma p \rightarrow \pi^{+} \pi^{0} n$ reaction,
the $\rho$ Kroll-Ruderman term ($T_{\rho {\rm KR}}$) and the $N^{*}$
terms dominate.
In this case,
the interference among $T_{\Delta {\rm KR}}$, $T_{\rho {\rm KR}}$
and $T_{N^{*}\rho N}$
is important and gives rise to the bump in the cross section.
So, we expect that the delicate balance of the interference is broken
in the nuclear medium by the collision broadening of the $\Delta$ and
the
$N^{*}$ and the pion distortion in the $\pi \Delta$ channel
and therefore the bump is
strongly damped due to the cooperative effect of the
broadening and the interference.

Now let us investigate the strong damping in the nuclear
photoabsorption.
For simplicity we use the Fermi gas model for a nucleus.
The absorption processes are assumed to consist of three components:
the quasi-free processes such as the one-pion photoproduction and
two-pion
photoproduction,
and genuine many-body absorption processes arising from the interaction
between the resonance (or pion) and the nucleon in a nucleus\cite{KMO}.

In the laboratory frame the cross section of
one-pion photoproduction off a proton in the nuclear matter is given by
\begin{eqnarray}
 \sigma^{\pi}_{p}& =&
\frac{1}{v}\frac{3Z}{8\pi({k^p_f})^3}\int^{k^p_f}_0
d\vec{p}_1\int\frac{d\vec{q}}{(2\pi)^3}
\frac{d\vec{p}}{(2\pi)^3}(2\pi)^4
\delta^4(k+p_1-q-p)  \nonumber \\
&\;&\times\frac{1}{2}\sum_{\lambda,\nu,\nu^{\prime}}
\sum_{t_{\pi},t_{N}}
|<\vec{q}t_{\pi}\vec{p}\nu^{\prime}t_{N}|T_{B}+T_{N^{*}}+T_{\Delta}|
\vec{k}\lambda\vec{p}_1\nu>|^2
\theta(|\vec{p}|-k^{t_{N}}_f)
\frac{M^2}{2k2\omega_{\vec{q}}E_{\vec{p}_1}
E_{\vec{p}}},
\end{eqnarray}
where $T_{N^{*}}$ and $T_{\Delta}$ represent the $N^{*}$ and $\Delta$
resonance terms, respectively and $T_{B}$ is the background term.
${\vec{k}}$, ${\vec{p}_{1}}$, ${\vec{q}}$ and ${\vec{p}}$ are the
momenta of the incident photon, target proton, outgoing pion
and outgoing nucleon, respectively.
$E_{\vec{p}_{1}}$, $\omega_{\vec{q}}$, and $E_{\vec{p}}$ are the
energies of the target proton, outgoing pion, and outgoing nucleon,
respectively. 
$Z$ and $v$ denote the proton number and the relative velocity between the 
photon and nucleus, respectively.
$k_{f}^{t_{N}}$ is the Fermi momentum depending on the isospin quantum
number $t_{N}$.
The notation for all other spin-isospin quantum numbers are
self-explanatory.
The $N^{*}$ resonance term is expressed as
\begin{eqnarray}
T_{N^*}&=&F_{\pi NN^*}
G_{N^{*}}(s)
\tilde{F}^{\dagger}_{\gamma PN^*},\\
G_{N^{*}}(s)&=&
               [\sqrt{s}-(M_{N^*}(s)+\delta M_{N^{*}})
                    +i(\Gamma_{N^*}(s)+\Gamma_{N^*sp})/2]^{-1},
\end{eqnarray}
where $\sqrt{s}$ is the total center of mass energy.
$M_{N^{*}}$ and $\Gamma _{N^{*}}$ in the $N^{*}$ propagator $G_{N^{*}}$
are the mass and the free width of $N^{*}$, respectively,
which are given so as to describe the energy dependence of the $\pi N$
$D_{13}$-wave scattering amplitude and the branching ratios at the
resonance energy.
$\delta M_{N^{*}}$ and $\Gamma _{N^{*}sp}$ are the mass shift and
spreading width due to the collisions between
$N^{*}$ and other nucleons, respectively.
$F_{\gamma P N^{*}}^{\dagger}$ and $F_{\pi N N^{*}}$ are the
$\gamma P N^{*}$ and $\pi N N^{*}$ vertex functions, respectively,
of which detail forms are given in Ref. \cite{ochi}.
The $\Delta$ resonance term $T_{\Delta}$ is written in a similar
form with $T_{N^{*}}$ but is important only at the low energy range
less than 500 MeV.
The Pauli blocking effect for the $\Delta$ decay into $\pi N$
becomes non-negligible at low energies, since the probability of the
nucleon
being emitted with a small momentum increases compared with the energy
region
of the $N^*$ resonance\cite{KMO,hirata}.
Thus, we include the Fock term in the $\Delta$ propagator
to modify the free $\Delta$ self-energy in addition to the collision
width.
The integration over final particle momenta in Eq. (1) is performed by
using variables defined in the $\gamma N$ center of mass system.
We will also calculate the cross section of two-pion photoproduction
in the same way.
%%%%%%%%%%%%%%%%%%%%%%%%%%%%%%%%%%%%%%%%%%%%%%%%%%%%%%%%%%%%%%%%%%%%%%

The cross section of the two-pion photoproduction off a proton is given
by
\begin{eqnarray}
\sigma^{2\pi}_p&=&
                  \frac{1}{v}\frac{3Z}{8\pi({k^p_f})^{3}}
\int^{{k^p_f}}_0d\vec{p}_1
\int\frac{d\vec{q}_1}{(2\pi)^3}
\frac{d\vec{q}_2}{(2\pi)^3}
\frac{d\vec{p}}{(2\pi)^3}
(2\pi)^4\delta^4(k+p_1-q_1-q_2-p)
\nonumber
 \\
&\;\;\;&\times
\frac{1}{2}\sum_{\lambda,\nu,\nu^{\prime}}
\sum_{t_{{\pi}_1}t_{{\pi}_2}t_N}
|<\vec{q}_1t_{{\pi}_1}\vec{q}_2t_{{\pi}_2}\vec{p}t_N\nu^{\prime}| T |
\vec{k}\lambda\vec{p}_1\nu>|^2
\theta(|\vec{p}|-k^{t_{N}}_f)
\frac{M^2}
{2k2\omega_{\vec{q}_1}2\omega_{\vec{q}_2}E_{\vec{p}_1}E_{\vec{p}}},
\end{eqnarray}
where ${\vec{q}_1}$ and ${\vec{q}_2}$ are the momenta of the outgoing
pions.
The medium-modified $T$ matrix is expressed as
\begin{eqnarray}
T&=&T_{\Delta KR}+ T_{\Delta PP}+ T^s_{N^*\pi\Delta}+ T^d_{N^*\pi\Delta}
    + T_{\rho {\rm KR}}+ T_{N^*\rho N}.
\end{eqnarray}
The $\Delta$ Kroll-Ruderman term is written as
\begin{eqnarray}
T_{\Delta KR}&=&
F_{\pi N\Delta}
G_{\pi \Delta}(s,\vec{p}_{\Delta})
F^{\dagger}_{\Delta KR},
\\
G_{\pi \Delta}(s,\vec{p}_{\Delta})&=&
        [\sqrt{s}-\omega_{\pi}(\vec{p}_{\Delta})
-(M_{\Delta}(s,\vec{p}_{\Delta})+\delta M_{\Delta})
+i(\Gamma_{\Delta}(s,\vec{p}_{\Delta})+\Gamma_{\Delta sp})/2
-V_{\pi}(\vec{q}_{\pi})]^{-1} ,
\end{eqnarray}
Here $F_{\pi N\Delta}$, 
is the $\pi N \Delta$ vertex function,
and $F^{\dagger}_{\Delta {\rm KR}}$ 
is the $\Delta$ Kroll-Ruderman vertex function.
The detail forms of vertex functions are given in Ref. \cite{ochi}.
$V_{\pi}(\vec{q}_{\pi})$ is the pion self-energy due to the
distortion.
$\vec{p}_{\Delta}$ is the $\gamma N$ center of mass momentum of $\Delta$
and
$\vec{q}_{\pi}$ is the outgoing pion momentum.
$M_{\Delta}$ and $\Gamma_{\Delta}$ in the propagator $G_{\pi \Delta}$
are the mass and the free width of $\Delta$, respectively
and $\delta M_{\Delta}$ and $\Gamma_{\Delta sp}$ are
the mass shift and collision width, respectively.
The other terms in the r.h.s. of Eq. (5) are expressed in a similar way.
The detail form of free $T$ matrices are given in Ref. \cite{ochi}.

In addition to the one- and two-pion photoproduction
processes,
there are three genuine many-body processes which are shown
in Figs. \ref{fig:fig1}(f), \ref{fig:fig1}(g) and \ref{fig:fig1}(h).
The cross section for Fig. \ref{fig:fig1}(f) is given by
\begin{eqnarray}
\sigma^p_{N^*(A-1)}&=&\frac{1}{v}\frac{3Z}{8\pi (k^p_{f})^{3}}
\int^{k^p_{f}}_{0}d\vec{p}_1\Gamma_{N^*sp}
\frac{1}{2}\sum_{\lambda\nu\nu_{N^*}}|G_{N^*}(s)
<\vec{p}_{N^*}\nu_{N^*}|\tilde{F}^{\dagger}_{\gamma PN^*}|\vec{k}\lambda
\vec{p}_1\nu>|^2\frac{1}{2k}\frac{M}{E_{\vec{p}_1}} .
\end{eqnarray}
The cross section for Fig. \ref{fig:fig1}(g) has a similar form with Eq.
(15).
The cross section for Fig. \ref{fig:fig1}(h) is given by
\begin{eqnarray}
\sigma^p_{\pi \Delta (A-1)}&=&\frac{1}{v}
                              \frac{3Z}{8 \pi{(k^p_{f})}^3}
\int^{k^p_{f}}_{0}
d\vec{p}_1\int\frac{d\vec{q}}{(2\pi)^3}
\frac{d\vec{p}}{(2\pi)^3}
(\Gamma_{\Delta sp}+2{\rm Im} V_{\pi}(\vec{q}))
(2\pi)^{3}\delta(\vec{k}+\vec{p}_{1}-\vec{q}-\vec{p})
\nonumber
\\
&\;\;&\times
\frac{1}{2}\sum_{\lambda\nu\nu_{\Delta}}|G_{\pi \Delta}(s,p_{\Delta})
<\vec{q}t_{\pi}\vec{p}\nu_{\Delta}t_{\Delta}|F^{\dagger}
_{\gamma P\pi \Delta}|\vec{k}\lambda \vec{p}_1\nu>|^2
\frac{1}{2k2\omega_{\vec{q}}}\frac{M}{E_{\vec{p}_1}} ,
\end{eqnarray}
where
$F_{\gamma P \pi \Delta}^{\dagger}$ describes the
$\gamma P \rightarrow \pi \Delta$ transition corresponding to
Figs. \ref{fig:fig1}(a)-\ref{fig:fig1}(d).
The cross sections of photoabsorption on a neutron in the nuclear matter
are also given in a similar form with those of a proton.

In the numerical calculations,
we make an approximation that the target nucleon's momentum
${\vec{p}_1}$
in the $T$-matrix elements is taken to be zero.
Although the effect of Fermi motion makes the bump of the cross section
small and broad, the missing strength of the cross section
cannot be explained at all with its effect alone\cite{phe2,phe3}.
Therefore it is clear that one has to seek the other possibilities to
understand the strong suppression.
We think the above approximation is sufficient to investigate the
qualitative
aspect of the damping mechanism.
In order to evaluate the cross sections, we need the values of the mass
shifts and collision widths of $\Delta$ and $N^{*}$, and the pion
self-energy $V_{\pi}$.
The mass shift and collision width of $\Delta$ have already known in the
studies of pion-nucleus scattering using $\Delta$-hole
model\cite{hirata,weise} where they can be identified to the spreading
potential.
The spreading potential found in these studies is almost
energy-independent.
We take $\delta M_{\Delta}=10$ MeV and $\Gamma_{\Delta sp}=80$ MeV.
As the pion self-energy, we adopt the pion optical potential used by
Arima {\it et al}.\cite{arima}.
$V_{\pi}$ has a strong energy-dependence because the pion-nucleon
interaction
in the relevant energy region is dominated by the $\Delta$ resonance.
As for the mass shift and collision width of $N^{*}$,
i.e., $\delta M_{N^{*}}$ and $\Gamma _{N^{*}sp}$,
there are no established values at present.
For simplicity, we assume that $\delta M_{N^{*}}$ and $\Gamma
_{N^{*}sp}$
are energy-independent like $\delta M_{\Delta}$ and $\Gamma _{\Delta
sp}$.
Furthermore, we choose these values so that the total nuclear
photoabsorption cross sections from 600 MeV to 800 MeV are reproduced.
We found 
$\delta M_{N^{*}}=-20$ MeV and $\Gamma _{N^{*} sp}=80$ MeV,
where the collision width of $N^{*}$ is the same as that of $\Delta$.
The calculated total photoabsorption cross sections per nucleon
(thick solid line) are shown
in Fig. \ref{fig:fig2}. In this calculation
we use the Fermi momentum $k_{f}$ obtained using the central nuclear
density and found that
the Pauli blocking effect is very small ($\leq$ 5\%) above 500 MeV.
We also show each contribution of the absorption processes:
the one-pion photoproduction (dashed line), the two-pion photoproduction
(dotted line), the many-body absorption through the $\Delta$-nucleus
state and $N^*$-nucleus state (dash-dotted line) corresponding to
Figs. \ref{fig:fig1}(f) and
\ref{fig:fig1}(g) and the many-body absorption through the $\pi
\Delta$-nucleus state
(long dashed line) corresponding to Fig. \ref{fig:fig1}(h).
For comparison, our theoretical result (thin solid line) for the total
photoabsorption cross
section on a proton is shown in Fig.3.

In the one-pion photoproduction,
the bump near the mass of the $N^*$ disappears by the effect of the
spreading
potential for the $N^*$. The cross sections of the two-pion
photoproduction
are quite suppressed by the cooperative effect
between the following medium corrections: the spreading potentials for
the $\Delta$ and the $N^{*}$, the pion distortion and the change of the
interference among the related reaction processes.
The cross sections of the
other many-body processes are almost flat in the energy range above 600
MeV and small.
As a consequence of these effects,
the excitation peak around the position of the $D_{13}$ resonance in
the total nuclear
photoabsorption cross sections disappear differently from the
hydrogen.
Extremely large collision width is not necessary for the $N^{*}$ resonance. 
Since the Fermi motion broadens the resonance peak and suppresses its value,
the inclusion of its effect in our calculation would probably lead to a
smaller collision width than the value obtained in our approach.

In order to confirm our conclusion and get more precise information
regarding the mass shift and the collision broadening for the $N^{*}$,
it is necessary to take into account the Fermi motion of the nucleon
in our calculation. This analysis is in progress.
Its results and the formalism used in this letter will be reported in
detail
elsewhere.

%%%%%%%%%%%%%%%%%%%%%%%%%%%%%%%%%%%%%%%%%%%%%%%%%%%%%%%%%%%%%%%%%%%%%%
\begin{figure}[h]
\caption{
The total photoabsorption cross section on a proton.
The dashed line is obtained by using SM95 amplitudes of
Arndt {\it et al.}\protect\cite{arndt}.
Solid and dash-dotted lines are theoretical predictions by our model
\protect\cite{ochi}.
Solid circles represent the data of total photoabsorption
cross section on a proton\protect\cite{exp5}.
Open squares represent the data of two-pion photoproduction cross section
\protect\cite{exp2}.
}
\label{fig:GP}
\end{figure}
\begin{figure}[h]
\caption{Diagrams for the two-pion photoproduction on a nucleon
and genuine many-body absorption processes on a nucleus.
(a) The $N^{*} \rightarrow \pi \Delta$ contribution.
(b) The $N^{*} \rightarrow \rho N$ contribution.
(c) The $\Delta$ Kroll-Ruderman term.
(d) The $\Delta$ pion-pole term.
(e) The $\rho$ Kroll-Ruderman term.
(f) The many-body absorption process through the $N^{*}$.
(g) The many-body absorption process through the $\Delta$.
(h) The many-body absorption process through the $\pi \Delta$.
A is the mass number of the target nucleus.
}
\label{fig:fig1}
\end{figure}
\begin{figure}[h]
\caption{The total nuclear photoabsorption cross section on nuclei.
The thick solid line is the full calculation. The dashed and dotted
lines are
the contributions of the one-pion and two-pion production, respectively.
The dash-dotted line is the contribution for the processes of Figs. 2(f)
and
2(g). The long dashed line is the contribution for the process of
Fig. 2(h).
The thin solid line is the theoretical result for the total cross section
on a proton calculated by our model.
%The thin solid line is the result by using our model of the total cross section
%on a proton.
Experimental data are taken from Ref. \protect\cite{exp3}}
\label{fig:fig2}
\end{figure}
%\onecol
\newpage
\pagestyle{empty}
\begin{figure}[H]
\leavevmode\psfig{file=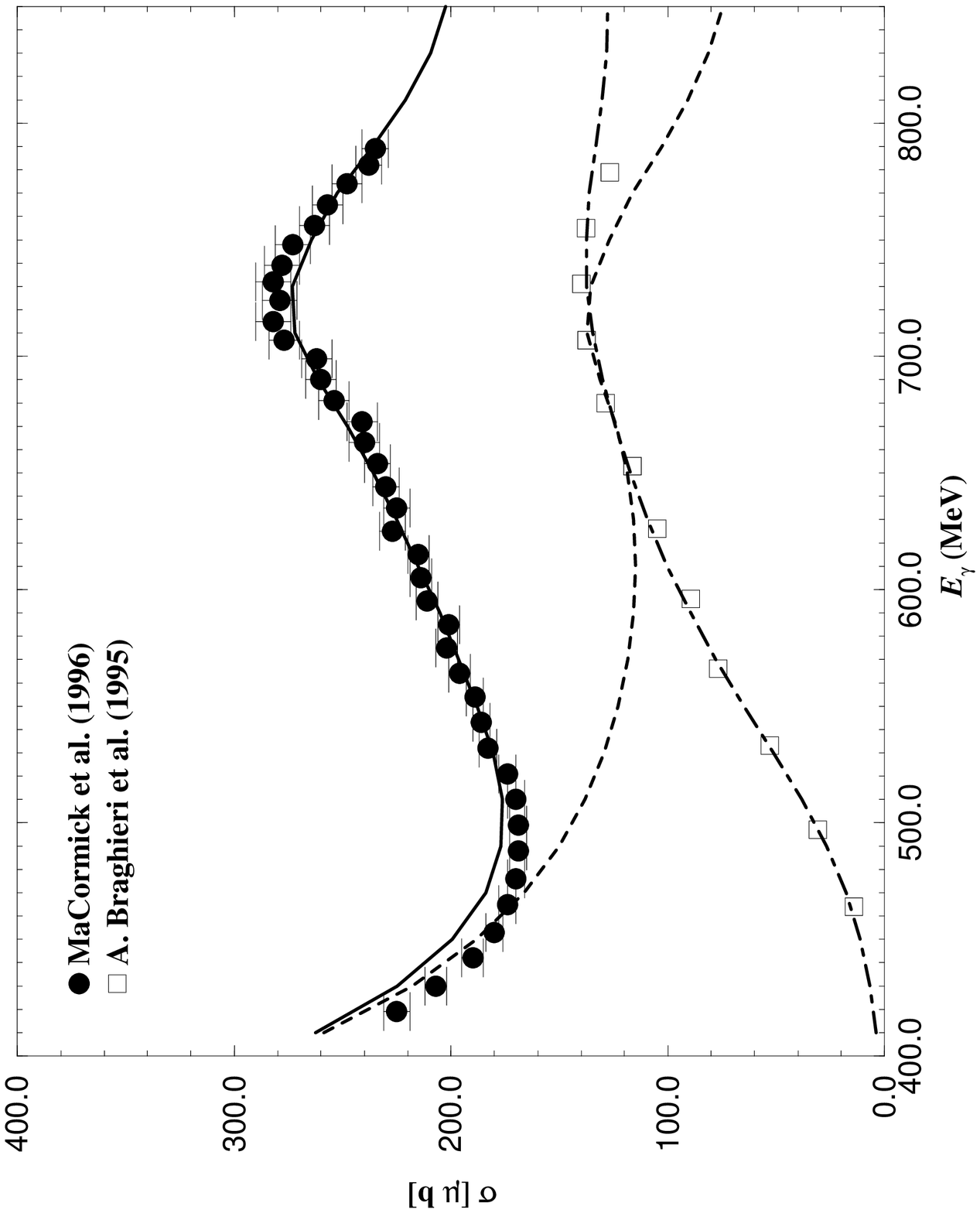,height=22cm,width=17cm}
\begin{center}
Fig. \ref{fig:GP}
\end{center}
\end{figure}
\begin{figure}[H]
\begin{center}
\begin{tabular}[h]{cc}
\leavevmode\psfig{file=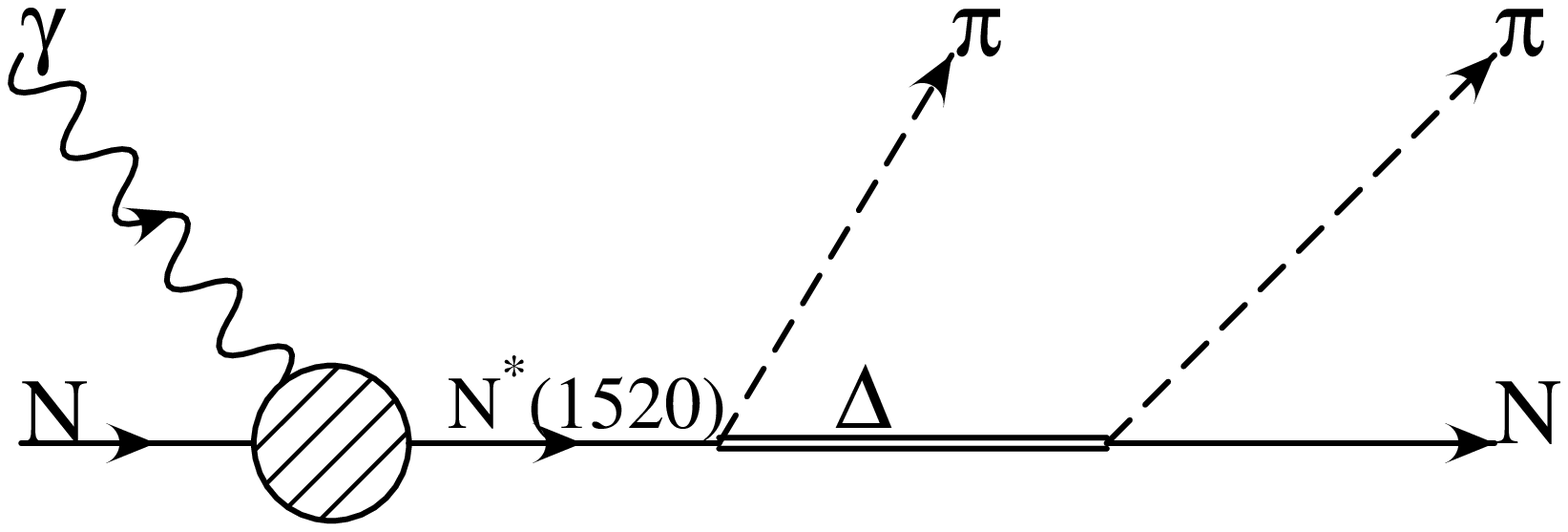,width=6cm}
&
\leavevmode\psfig{file=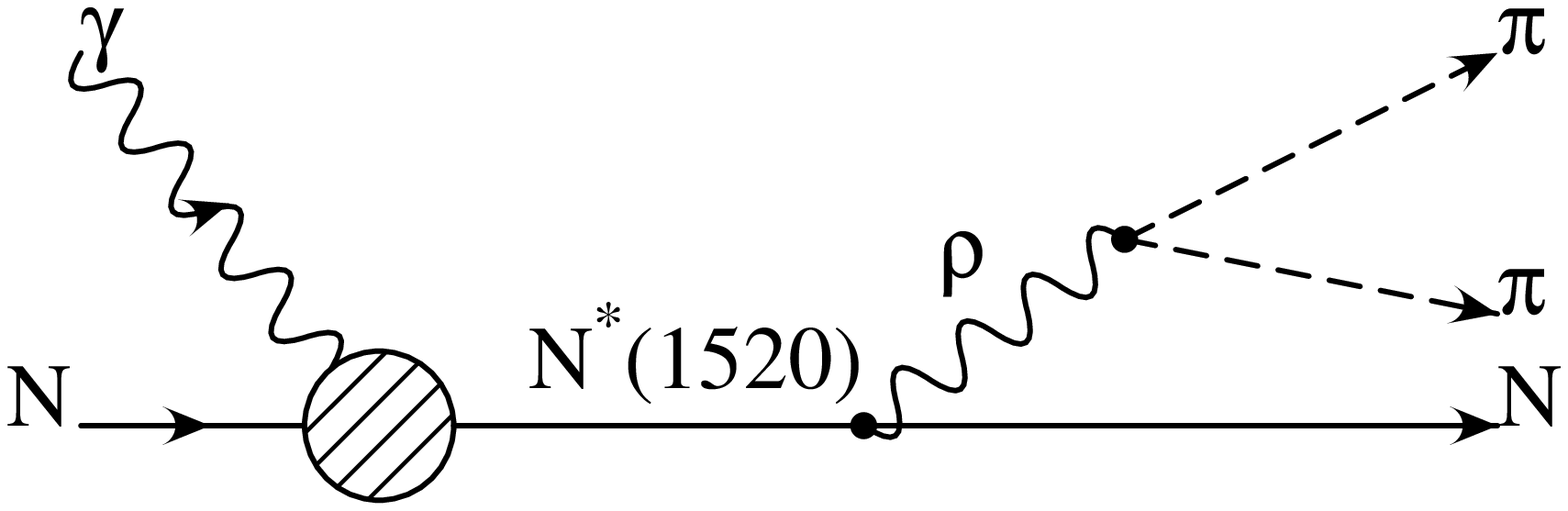,width=6cm}
\\
(a) & (b)
\\
\leavevmode\psfig{file=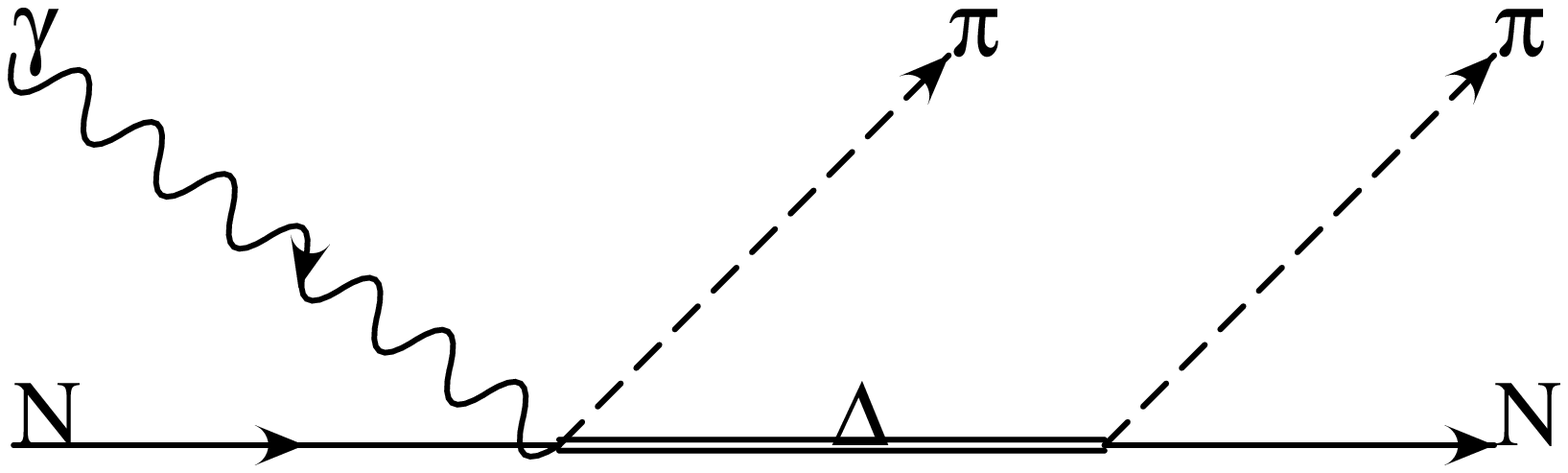,width=6cm}
&
\leavevmode\psfig{file=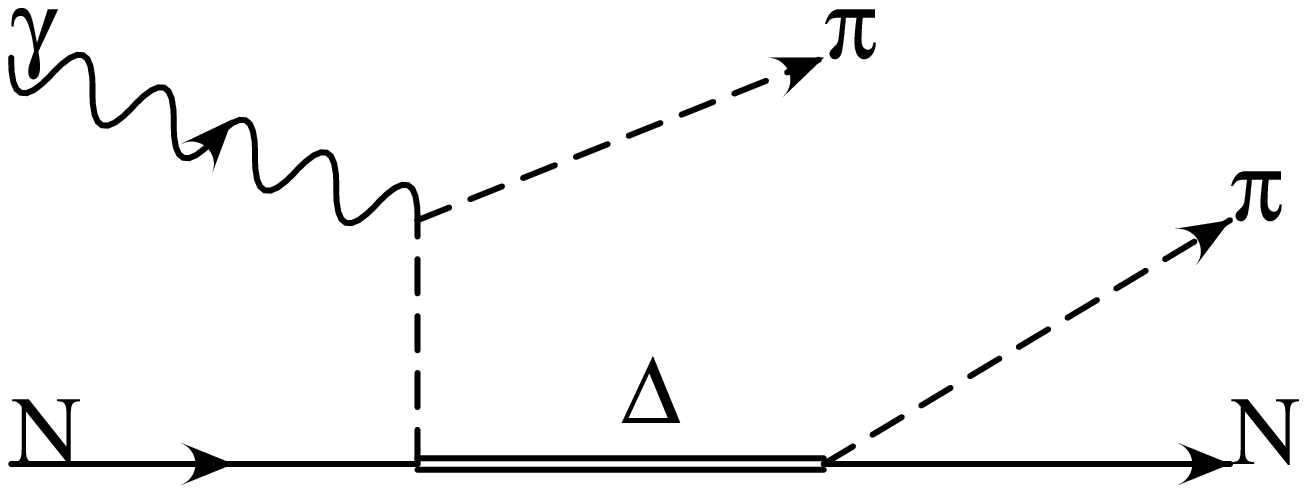,width=6cm}
\\
(c) & (d)
\\
\leavevmode\psfig{file=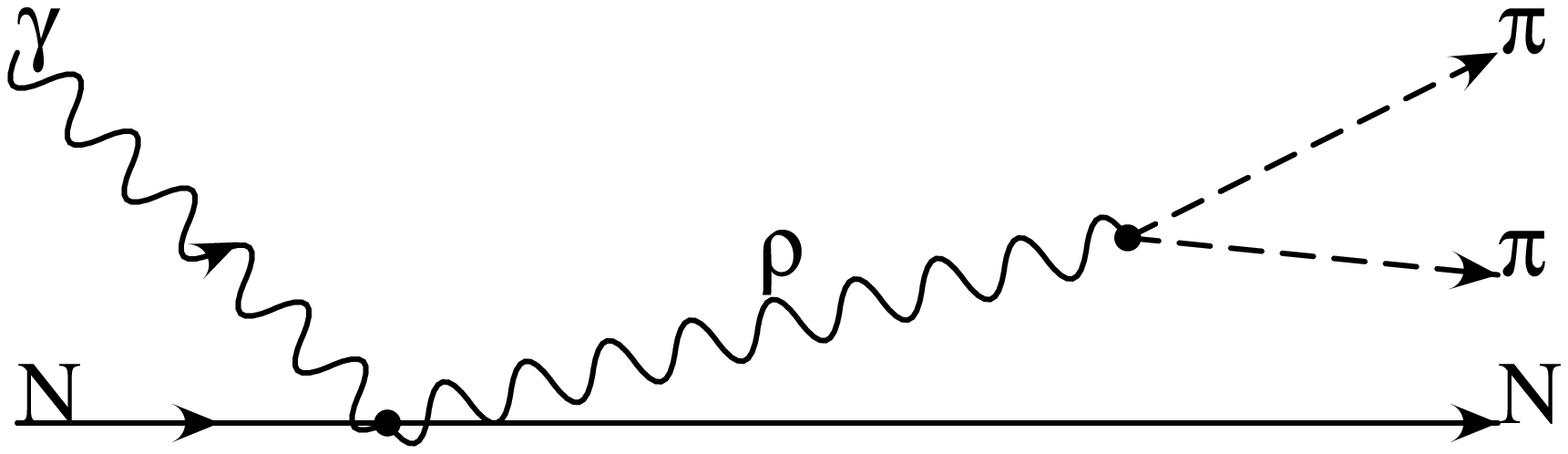,width=6cm}
&
\leavevmode\psfig{file=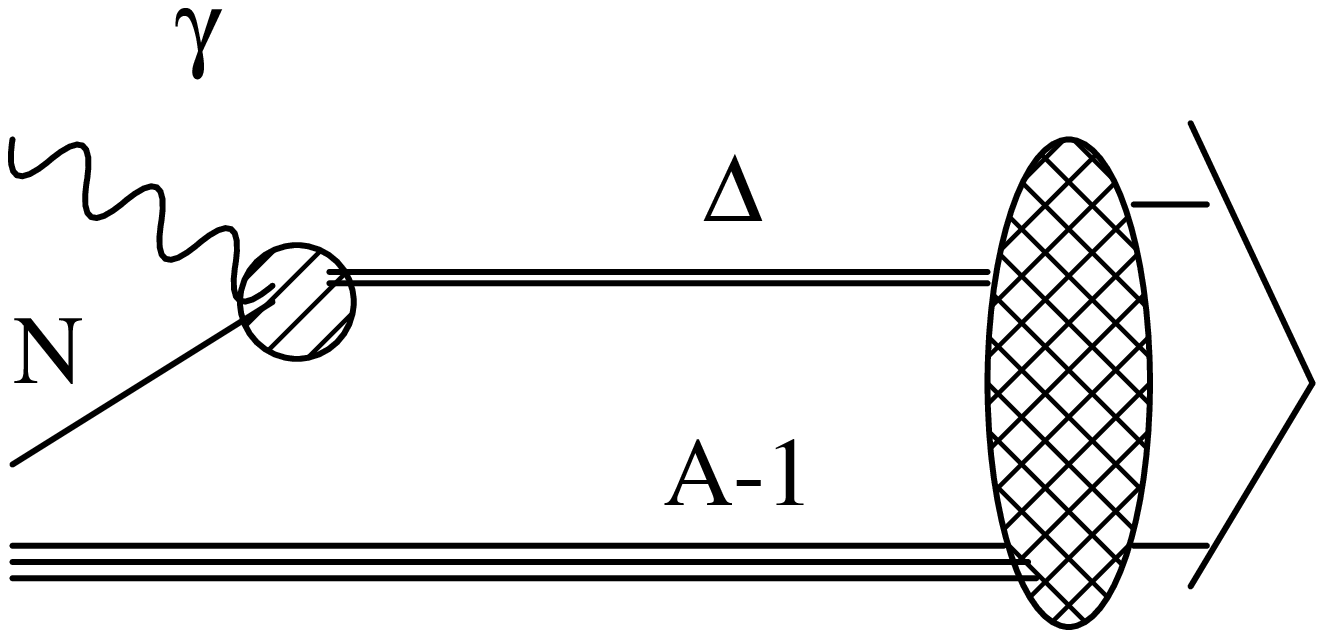,width=6cm}
\\
(e) & (f)
\\
\leavevmode\psfig{file=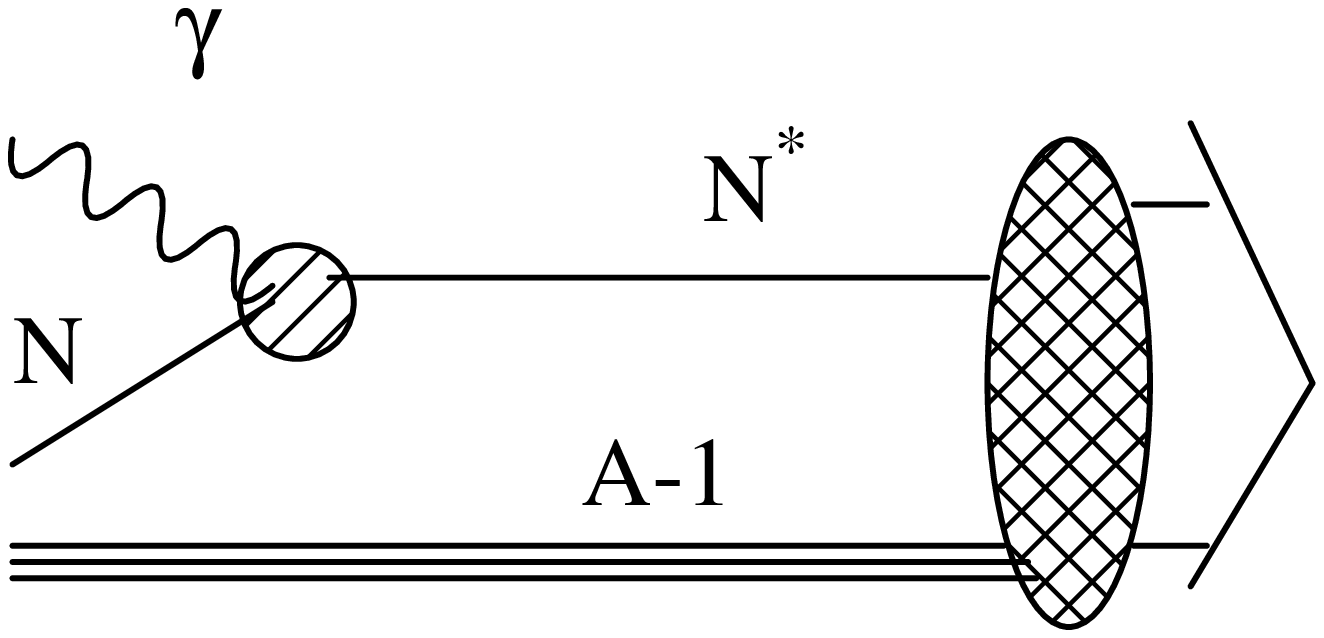,width=6cm}
&
\leavevmode\psfig{file=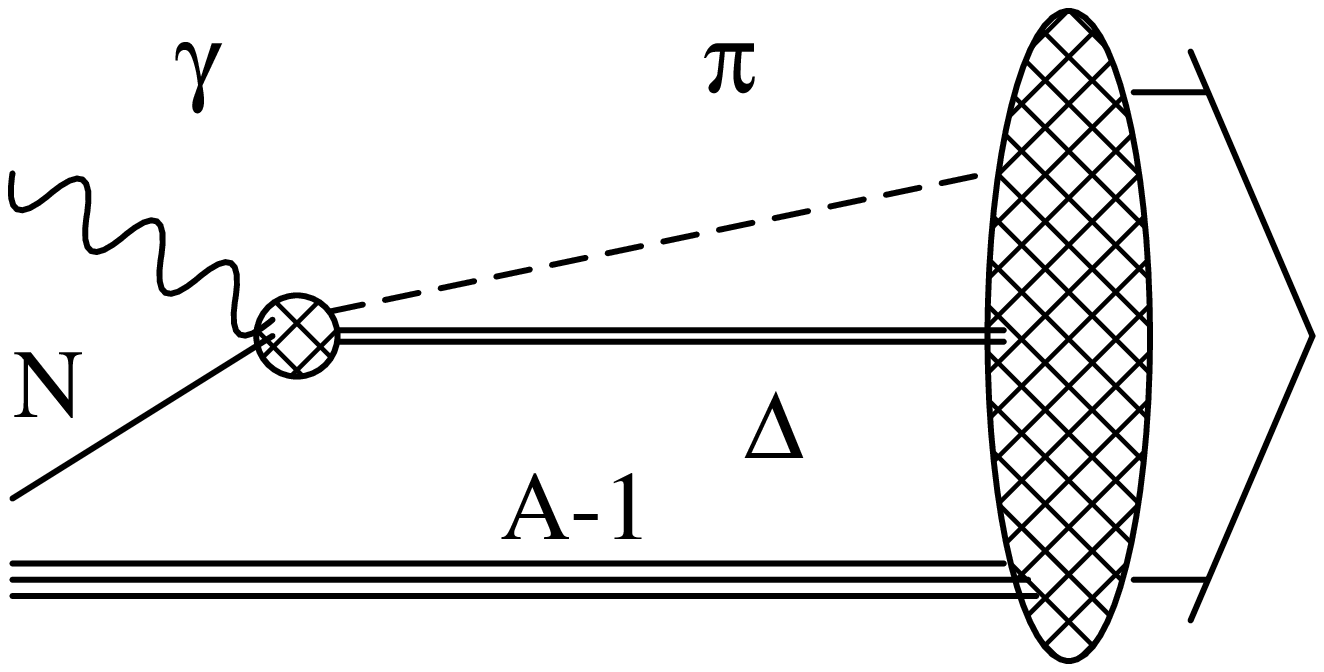,width=6cm}
\\
(g) & (h)
\end{tabular}
\end{center}
\end{figure}
\begin{center}
Fig. \ref{fig:fig1}
\end{center}
\newpage
\begin{figure}[H]
\leavevmode\psfig{file=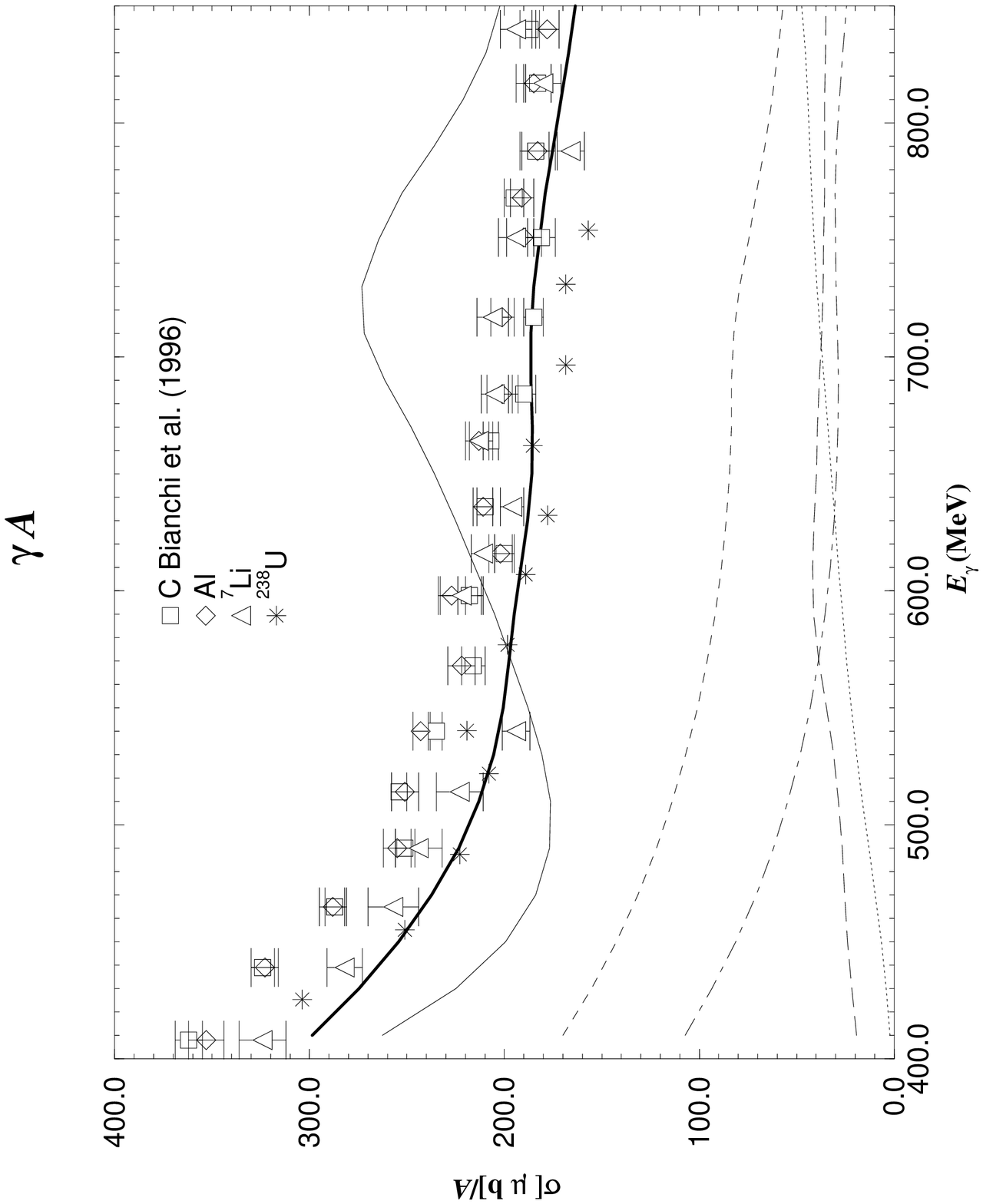,height=22cm,width=17cm}
\end{figure}
\begin{center}
Fig. \ref{fig:fig2}
\end{center}
%%%%%%%%%%%%%%%%%%%%%%%%%%%%%%%%%%%%%%%%%%%%%%%%%%%%%%%%%%%%%%%%%%%%%%
\end{document}